\def\be{\begin{equation}}
\def\ee{\end{equation}}
\def\bea{\begin{eqnarray}}
\def\eea{\end{eqnarray}}
\def\a{\alpha}
\def\b{\beta}

\def\r{\rho}
\def\t{\theta}

\documentclass[12pt]{article}

\usepackage{amsmath, amsgen,amstext,amsbsy,amsopn,amsthm, amssymb }
\begin{document}

\title{The odd eight-vertex model}

\author{F. Y. Wu$^1$ and H. Kunz$^2$\\
$^1$Department of Physics, Northeastern University \\ Boston, 
Massachusetts 02115, USA \\
 $^2$ Institut de Physique Th\'eorique, Ecole Polytechnique F\'ed\'erale\\
Laussane, Switzerland}

\maketitle
\begin{abstract} 
We consider a vertex model on the simple-quartic lattice defined by line graphs on
the lattice for which there is always
an odd number
of lines incident at a vertex.  This is the odd 8-vertex model which has
eight possible vertex  configurations.  We establish that the odd
 8-vertex model is equivalent to a staggered 8-vertex model. Using 
this equivalence we
 deduce the solution of the odd 8-vertex model when the weights satisfy a 
free-fermion condition. 
It is found that the free-fermion model exhibits no phase transitions
in the regime of positive vertex weights.
We also establish the complete equivalence of the free-fermion odd 8-vertex model
 with the free-fermion 8-vertex model solved by Fan and Wu.
Our analysis leads to several Ising model representations of the free-fermion model
with pure 2-spin interactions.

\end{abstract}
 
\newpage
\section{Introduction}
In a seminal work which opened the door to  a new era of exactly solvable
models in statistical mechanics, Lieb  \cite{lieb67,lieb671} in 1967
 solved the problem of the residual entropy 
of the square ice.
 His work led soon thereafter to the solution of a host
of  more general lattice models
of phase transitions.  These
include the five-vertex model \cite{wu67, wu96},
the F model \cite{lieb672}, the KDP model \cite{lieb673},
the general six-vertex model \cite{sy67}, the free-fermion model solved by
Fan and Wu \cite{fanwu}, and  the 
symmetric 8-vertex model solved by Baxter \cite{baxter72}.
All these previously considered models are  described by line graphs drawn on a simple-quartic lattice 
where the number of lines incident at each vertex is  even,
and therefore can be regarded as the ``even" vertex models.

\medskip
 In this paper we consider the {\it odd} vertex models, a problem that
does not to have  attracted much past  attention.  
 Again, one draws line graphs on the simple-quartic lattice but with the restriction
that the number of lines incident at a vertex is always odd.
 There are again eight possible ways of drawing lines at
a vertex, and  this leads to the {\it odd 8-vertex model}.
 Besides being a challenging  mathematical problem by itself, as we shall see
the odd 8-vertex model includes some well-known unsolved lattice-statistical problems.
It also finds applications in
 enumerating  dimer configurations \cite{wu03}.

 \medskip
 Consider a simple-quartic lattice of $N$ vertices and draw lines on the lattice
such that the number of lines incident at a vertex
 is always odd, namely, 1 or 3.
There are eight possible 
vertex configurations are shown in Fig. 1.  To  vertices of type $i$ $(=1,2,\cdots,8)$
we associate weights $u_i> 0$.
 Our goal is to compute the partition function
\be
Z_{12\cdots 8} \equiv Z(u_1,u_2,\cdots,u_8) 
  = \sum_{\rm o.l.g.} u_1^{n_1}u_2^{n_2}\cdots u_8^{n_8} \label{part}
\ee
where the summation is taken over all aforementioned odd line graphs,
and $n_i$ is the number of
vertices of the type ($i$).
The per-site ``free energy" 
is then computed as
\be
f = 
 \lim_{N\to \infty} \frac 1 {N} \ln Z_{12\cdots 8}.
\ee

The partition function (\ref{part}) possesses obvious symmetries. 
An  edge can either have a line is or be vacant.  
  By reversing the line-vacancy role  
 one obtains the symmetry
\be
Z_{12345678} = Z_{21436587}.  \label{symmetry1}
\ee
Similarly, the left-right
and up-down symmetries dictate the equivalences
\be
Z_{12345678} = Z_{12347856} = Z_{34125678}, \label{symmetry2}
\ee
and successive $90^o$ counter-clockwise rotations of the lattice lead to 
\be
Z_{12345678} = Z_{78561243}=Z_{34127856} = Z_{56783421}. \label{symmetry3}
\ee
These are intrinsic symmetries of the odd 8-vertex model.

\medskip
The odd 8-vertex model encompasses an unsolved
Ashkin-Teller model \cite{ashkinteller} as a special case (see below).
 It also generates other known solutions.
For example, it is clear from Fig. 1 
that by  taking
\bea
&& u_1=y,\quad \quad \quad u_3=1 \nonumber \\
&& u_5=x, \quad \quad \quad u_7=1 \nonumber \\
&&  u_2=u_4 =u_6 = u_8 =0 \label{dimer}
\eea
(and assuming periodic boundary conditions)
 the line graph  generate close-packed
dimer configurations  on the simple-quartic lattice with activities $x$ and $y$.
The solution of (\ref{part}) in this case
is  well-known \cite{fishertemperley, kasteleyn}.

\section{Equivalence with a staggered vertex model}
 Our approach to the odd 8-vertex model is to 
explore its equivalence with a staggered 
8-vertex model.  
   We first recall the definition
of a staggered 8-vertex model \cite{hsue}. 
 
\medskip
A staggered 8-vertex model  is an (even) 8-vertex model with
sublattice-dependent vertex weights.
It is   defined by 16 vertex
weights $\{\omega_i\}$ and $\{\omega_i'\}$, $i=1,2\cdots,8$,
one for each sublattice, associated with the 8
(even) line graph configurations shown in Fig. 2.
 
\medskip
The partition function of the staggered 8-vertex model is
\be
Z_{\rm stag} (\omega_1,\omega_2,\cdots,\omega_8;\> \omega'_1,\omega'_2,\cdots,\omega'_8) 
= \sum_{\rm e.l.g.}\prod_{i=1}^8 \big[ {\omega_i}^{n_i}({\omega'_i})^{n'_i} \big] \label{spart}
\ee
where the summation is taken over all even line graphs, and $n_i$ and $n'_i$ are,
respectively, the numbers of vertices with weights $\omega_i$ and $\omega_i$.
It is convenient to abbreviate the partition function by writing
\be
Z_{\rm stag} (\omega_1,\omega_2,\cdots,\omega_8;\> \omega'_1,\omega'_2,\cdots,\omega'_8)
\equiv Z_{\rm stag}(12345678;\> 1'2'3'4'5'6'7'8') . \label{16part}
\ee

When $\omega_i=\omega'_i$ for all $i$, the staggered 8-vertex model
reduces to the usual 8-vertex model with  uniform weights, which
remains unsolved for general $\omega_i$.  
When $\omega_i\not=\omega'_i$ the problem is obviously even harder.
 The consideration of  the sublattice symmetry implies that we have
\be
Z_{\rm stag} (12345678;\> 1'2'3'4'5'6'7'8') =
Z_{\rm stag} (1'2'3'4'5'6'7'8';\> 12345678). \label{sub1}
\ee 

Returning to the odd 8-vertex model we have the following result: 

\medskip
\noindent
{\it Theorem: The odd 8-vertex model (\ref{part}) is equivalent to
a staggered 8-vertex model (\ref{16part}) with the equivalence 
  \bea
Z_{12\cdots 8} &=& Z_{\rm stag}(u_1,u_2,u_3,u_4,u_5,
u_6,u_7,u_8;\ u_3, u_4, u_1,u_2,u_8,u_7,u_6,u_5) \nonumber \\
   &=&  Z_{\rm stag}(u_5,u_6,u_8,u_7,u_1,u_2,u_3,u_4;\  
u_7,u_8,u_6,u_5,u_4,u_3,u_1,u_2), \nonumber
\eea
or, in abbreviations,}
   \bea
Z_{12\cdots 8} &=& Z_{\rm stag}(12345678;\> 34128765) \nonumber \\
   &=&  Z_{\rm stag}(56871243;\> 78654312). \label{theorem}
\eea

 \medskip
{\it Proof}:
Let  $A$ and $B$ be the two sublattices each having $N/2$ sites.
Consider the set $S$ of $N/2$ edges each of which connecting
an $A$ site to a $B$ site immediately below it. 
 By reversing  the roles of occupation and  vacancy on these edges,
 the vertex configurations of Fig. 1 are converted  into
  configurations with an even number of incident lines.  
   Because of the particular choice of $S$,  however,
  the vertex weights are sublattice-dependent and we have 
 a  {\it staggered} 8-vertex model.

\medskip
For sites on sublattice $A$, the conversion 
  maps  a vertex   type $(i)$ in Fig. 1
 into a type $(i)$  
in Fig. 2   so that $\omega_i=u_i$  for {\it all} $i$ on $A$.
At $B$ sites the conversion maps type (3) in
Fig. 1 to type (1) in Fig. 2, (4)  to (2) with $\omega_1'=u_3, \ \omega_2'=u_4$, etc.
   Writing compactly and rearranging the $B$ weights according to configurations
in Fig. 2, 
the mappings are
\bea
&&\omega\{12345678\}\ \to u\{12345678\}, \hskip 1cm {\rm at\ } A{\rm \ sites} \nonumber \\ 
&&\omega'\{12345678\} \to u\{34128765\}, \hskip 1cm {\rm at \ }B{\rm \ sites}.
\eea 
This establishes the first line in (\ref{theorem}).

\medskip
The line-vacancy conversion can also be carried out for  any of the three
other edge sets connecting every $A$ site to the $B$
site above it, on the right, or on the left.
 It is readily verified that these considerations lead to the equivalence
given by the second line in (\ref{theorem}),
 and two others obtained from (\ref{theorem}) by applying
 the sublattice symmetry (\ref{sub1}).  Q.E.D.

\medskip
{\it Remark}: Further  equivalences can be obtained by combining 
(\ref{symmetry1}) - (\ref{symmetry3}) with the sublattice symmetry (\ref{sub1}).

\medskip
The special case of
\bea
&& u_1=u_2=u_3=u_4 \nonumber \\
&& u_5=u_6, \hskip .45cm u_7=u_8
\eea
is an Ashkin-Teller model as formulated  in \cite{wulin} which remains unsolved.
Another special case is when the  weights satisfy 
\be
u_1u_2 +u_3u_4 =u_5u_6 +u_7u_8 .\label{ff}
\ee
Then from (\ref{theorem}) the staggered 8-vertex model weights satisfy
 the free-fermion condition 
\bea
\omega_1\omega_2 +\omega_3\omega_4 &=& \omega_5\omega_6 +\omega_7\omega_8\ \nonumber \\
\omega_1'\omega_2' +\omega_3'\omega_4'&=&\omega_5'\omega_6' +\omega_7'\omega_8'\  \label{ff1}
\eea
for which the solution has been obtained in \cite{hsue}.
 This case is discussed  in the next section.

\section{The free-fermion solution}
In this section we consider the  odd 8-vertex model
(\ref{part}) satisfying the free-fermion condition (\ref{ff}).
In the language of the first line of the equivalence (\ref{theorem})
we have the staggered vertex weights
 \bea
&&\omega_1=\omega_3' = u_1, \hskip 1.5cm  \omega_2=\omega_4'= u_2 \nonumber \\
&&\omega_3=\omega_1' = u_3, \hskip 1.5cm  \omega_4=\omega_2'= u_4 \nonumber \\
&&\omega_5=\omega_7' = u_5, \hskip 1.5cm  \omega_6=\omega_8'= u_6 \nonumber \\
&&\omega_7=\omega_5' = u_1, \hskip 1.5cm  \omega_8=\omega_6'= u_8 , \label{Bweights}
\eea
and hence the condition (\ref{ff1}) is satisfied.  This leads to the free-fermion 
staggered 8-vertex model studied
in \cite{hsue}.  Using results of \cite{hsue} and the weights (\ref{Bweights}), 
 we obtain after a little reduction the solution
  \be
f = \frac 1 {16\pi^2} \int_0^{2\pi} d\t\int_0^{2\pi} d\phi \ln F(\t, \phi) 
\label{fffe}
\ee
  where
\be
 F(\t, \phi) = 2 A +2D \cos(\t-\phi) +2E \cos (\t +\phi)
  +4\Delta_1 \sin^2\phi
+4\Delta_2 \sin^2\t \nonumber
 \ee
with 
\bea
A&=& (u_1u_3+u_2u_4)^2+(u_5u_7+u_6u_8)^2 \nonumber \\
D&=& (u_5u_7)^2 +(u_6u_8)^2 - 2u_1u_2u_3u_4  \nonumber \\
E&=& -(u_1u_3)^2 -(u_2u_4)^2 +2u_5u_6u_7u_8  \nonumber \\
\Delta_1 &=& (u_1u_2-u_5u_6)^2 > 0\nonumber \\
\Delta_2 &=& (u_3u_4-u_5u_6)^2 > 0 .\label{ade}
\eea
As an example, specializing (\ref{fffe}) to the weights (\ref{dimer}) for the dimer
problem,
we have $A=x^2+y^2$,\ $D=x^2,\ E=-y^2$,\ $\Delta_1=\Delta_2=0$, and 
(\ref{fffe}) leads to the known dimer solution \cite{fishertemperley,kasteleyn}
\be
f_{\rm dimer} 
 =\frac 1 {\pi^2} \int_0^{\pi/2} d\omega\int_0^{\pi/2} d\omega'\ 
\ln (4 x^2 \sin ^2 \omega + 4 y^2 \sin^2 \omega'),
\ee
which has no phase transitions.  More generally for
  $u_i>0$ we have $A > |D| + |E|$  and hence 
\bea
F(\t ,\phi) >0.\nonumber
\eea
  As a result, 
the free energy $f$ given by (\ref{fffe}) is analytic and 
 there is no  singularity in $f$ implying that
the odd 8-vertex model has no phase transition.

\section{Equivalence with the free-fermion model of Fan and Wu}
The free energy (\ref{fffe}) is of  the form of that of the free-fermion 
model solved by Fan and Wu \cite{fanwu}. 
To see this we change
  integration variables in (\ref{fffe}) to
\be
\a=\t+\phi, \hskip 1cm \b=\t-\phi,
\ee
the expression (\ref{fffe}) then assumes the form
 \bea
f &=& \frac 1 {16\pi^2} \int_0^{2\pi} d\a\int_0^{2\pi} d\b \ln 
\big[ 2 A_1 +2E \cos\a +2D \cos \b \nonumber \\
 && \hskip .5cm -2\Delta_1 \cos(\a-\b) -2\Delta_2 \cos (\a +\b)\big] \label{fffe1}
\eea
where, after making use of (\ref{ff}),
\bea
A_1&=&  A+\Delta_1 +\Delta_2 \nonumber  \\ 
&=&(u_1u_2+u_3u_4)^2+(u_1u_3)^2+(u_2u_4)^2+(u_5u_7)^2+(u_6u_8)^2. \nonumber 
 \eea
Comparing (\ref{fffe1}) with
Eq. (16) of \cite{fanwu}, we find
 \be
f =  f_{\rm FF}/2
\ee
where $f_{\rm FF}$ is the per-site free energy of an 8-vertex model with
{\it uniform} weights $w_1, w_2, ..., w_8$
satisfying the free-fermion condition 
\be
w_1w_2+w_3w_4=w_5w_6+w_7w_8  \label{ff2}
\ee
and
 \bea
A_1&=&  \big(w_1^2+w_2^2+w_3^2+w_4^2\big)/2 \nonumber \\
D&=& w_1w_4- w_2w_3 \nonumber \\
E&=& w_1w_3- w_2w_4 \nonumber \\
\Delta_1&=& w_1w_2- w_5w_6 \nonumber \\
\Delta_2&=&w_5w_6- w_3w_4 \ .  \label{relation}
\eea
We can solve for $w_1,
w_2,w_3,w_4$ and $w_5w_6$ from the five equations in (\ref{relation}),
and  then determine $w_7w_8$ from (\ref{ff2}).
 
\medskip
By equating (\ref{relation}) with (\ref{ade}), it can be verified that one has
\bea
 (-w_1+w_2+w_3+w_4)^2 &=&2(A_1-D-E-\Delta_1 -\Delta_2) = v_1^2 \nonumber \\
(w_1-w_2+w_3+w_4)^2&=& 2(A_1+D+E-\Delta_1 -\Delta_2) = v_2^2\nonumber \\
 (w_1+w_2-w_3+w_4)^2 &=&2(A_1+D-E+\Delta_1 +\Delta_2) = v_3^2 \nonumber \\
 (w_1+w_2+w_3-w_4)^2 &=& 2(A_1-D+E+\Delta_1 +\Delta_2) = v_4^2 ,\label{relation1}
\eea
where\footnote{The apparent asymmetry in the expression of $v_3$ can be
traced to the choice of the edge set $S$ used in section 2 in deducing the
equivalent staggered 8-vertex model.} 
\bea
&& v_1= 2(u_1u_3+u_2u_4) \nonumber \\
&& v_2= 2(u_5u_7+u_6u_8) \nonumber \\
&& v_3= 2\sqrt{(u_1u_2+u_3u_4)^2 +(u_1u_3-u_2u_4)^2 +(u_5u_7-u_6u_8)^2}  \nonumber \\
&& v_4= 2(u_1u_2+u_3u_4) \label{v}.
\eea
Then, taking the square root of (\ref{relation1}), one obtains the explicit solution
\be
w_i = (v_1+v_2+v_3+v_4 - 2v_i)/4, \hskip 1cm i=1,2,3,4.
\ee
The 4th line of (\ref{relation}) now yields
\be
w_5w_6= w_1w_2-(u_1u_2 -u_5u_6)^2,
\ee
and $w_7w_8$ is obtained from (\ref{ff2}).

\medskip
The free-fermion model is known \cite{fanwu} to be critical at
\be
2w_i = w_1+w_2+w_3+w_4 , \hskip 1cm i=1,2,3,4. \label{ffcritical}
\ee
which is equivalent to $v_i=0$.  It is then clear from (\ref{v}) that 
the critical point (\ref{ffcritical}) lies outside the region $u_i >0$
and this confirms our earlier conclusion that the odd 8-vertex model does 
not exhibit a transition in the regime of positive weights.  Our results also
sown that the model with some $u_i = 0$, e.g., $u_7=u_8=0$, is
critical. This is reminiscent to the known fact of  the even vertex models that
the 8-vertex model 
is critical in the 6-vertex model subspace.

\section{Ising representations of the free-fermion model}
The free-fermion
odd 8-vertex model can be formulated as  Ising models with pure 2-spin
interactions in several different ways. 
 In the preceeding section we have established its equivalence
with the Fan-Wu free-fermion model.  Baxter \cite{baxter86} has shown that
the Fan-Wu free-fermion model is equivalent to
a checkerboard Ising model and that asymptotically it can be decomposed
into four overlapping Ising models.  It follows that the odd 8-vertex
model possesses the same properties, namely, it
is equivalent to a checkerboard Ising model and can be similarly 
decomposed asymptotically. We refer to \cite{baxter86} for details of analysis.
    
\medskip
An alternate Ising representation can be constructed as follows:
Consider  the equivalent staggered 8-vertex model
given in the first line of (\ref{theorem}).  
We place Ising spins on  dual lattice sites as shown in Fig. 2
and write the partition function as
\be
Z_{\rm Ising} = \sum_{\rm spin\>config.} \prod_{\rm A} W(a,b,c,d)\prod_{\rm B} W'(a,b,c,d)
\label{bfactor}
\ee 
where the summation is taken over all spin configurations, and $W$ and $W'$
are, respectively, the Ising Boltzmann factors associated with four spins
$a,b,c,d = \pm 1$ surrounding each $A$ and $B$ sites. Since the vertex 
to spin configuration mapping is $1:2$,  we have the equivalence
\be
Z_{12\cdots 8} =   Z_{\rm Ising}/2.
\ee

We next require the Ising Boltzmann factors $W$ and $W'$ to reproduce 
the vertex weights $\omega$ and $\omega'$ in (\ref{theorem}).
 Now to each vertex in the free-fermion model 
there are six independent parameters  after taking
into account the free-fermion condition (\ref{ff}) an overall constant.
We therefore need six Ising parameters which we introduce as
shown in Fig. 3 for $W(a,b,c,d)$ on sublattice $A$.  
Namely, we write
 \be
W(a,b,c,d) = 2 \r\ e^{M(ad-bc)/2 +P(cd-ab)/2} \cosh (J_1a+J_2b+J_3c+J_4d) \label{ising}
\ee
where $\r$ is an overall constant.
Explicitly, a perusal of Fig. 2 leads to the expressions
 \bea
u_1 &=& 2\r \cosh (J_1+J_2+J_3+J_4), \hskip 1.33cm
u_2 = 2\r \cosh (J_1-J_2+J_3-J_4)  \nonumber \\
u_3 &=& 2\r \cosh (J_1-J_2-J_3+J_4), \hskip 1.33cm
u_4 = 2\r \cosh (J_1+J_2-J_3-J_4)  \nonumber \\
u_5 &=& 2\r\ e^{M+P} \cosh (J_1-J_2+J_3+J_4), \ \ 
u_6 = 2\r\ e^{-M-P}\cosh (J_1+J_2+J_3-J_4)  \nonumber \\
u_7 &=& 2\r\ e^{P-M} \cosh (-J_1+J_2+J_3+J_4), \ \ 
u_8 = 2\r\ e^{M-P} \cosh (J_1+J_2-J_3+J_4) .\nonumber \\
\label{weights}
\eea
These weights satisfy the free-fermion condition (\ref{ff}) 
automatically.\footnote{Expressions in
Eq. (\ref{weights}) are the same as Eq. (2.5) in  \cite{baxter86}
except the
interchange of expressions  $u_7$ and $u_8$ due to the different ordering of
 configurations (7) and (8).}
 
\medskip
 Equation (\ref{weights}) can be used to solve for  $J_1,J_2,J_3,J_4,M,P$
and the overall constant $ \r$
in terms of the weights $u_i$.  
First, using the first four equations one solves for $J_1, J_2, J_3,J_4$ 
in terms of $\cosh^{-1} (u_i/2\r), i=1,2,3,4$. Then the overall constant
$\r$ is solved from the equation
\be
\frac{u_5u_6}{u_7u_8} = \frac{\cosh 2(J_1+J_3)+\cosh 2(J_2-J_4)}
{\cosh 2(J_1-J_3)+\cosh 2(J_2+J_4)}
\ee
and $M$, $P$ are given by
  \bea
e^{4M} &=&\Big(\frac{u_5u_8 }{u_6u_7}\Big)\Bigg[\frac{\cosh 2(J_1-J_4)+\cosh 2(J_2+J_3)}
{\cosh 2(J_1+J_4)+\cosh 2(J_2-J_3)}\Bigg],  \nonumber \\
e^{4P} &=&\Big(\frac{u_5u_7 }{u_6u_8}\Big)
\Bigg[\frac{\cosh 2(J_1-J_2)+\cosh 2(J_3+J_4)}
{\cosh 2(J_1+J_2)+\cosh 2(J_3-J_4)}\Bigg].
\eea

 \medskip
For $B$ sites, we note that the weights are precisely
those of $A$ sites with the interchanges $u_1 \leftrightarrow u_3,$\ 
$u_2 \leftrightarrow u_4,$\ $u_5 \leftrightarrow u_8,$\ $u_6 \leftrightarrow u_7$.
In terms of the spin configurations, these interchanges correspond to
 the negation of the spins $b$ and $c$.
 Thus we have
\bea
W'(a,b,c,d) &=&  W(a,-b,-c,d)  \nonumber \\ 
&=& 2 \r\ e^{M(ad-bc)/2 - P(cd-ab)/2} \cosh (J_1a-J_2b-J_3c+J_4d). \nonumber \\
\label{isingb}
\eea
This  Boltzmann factor is the same as (\ref{bfactor})
with the same $J_1,J_4,M, \r$ and the negation of $J_2, J_3,$ and $P$.
Namely, we have
\bea
&& J_1'=J_1, \quad J_2'=-J_2, \quad M' = M , \quad \r'=\r \nonumber\\
&& J_3'=-J_3, \quad J_4'=J_4, \quad P' = -P
\eea
 
 Putting the Ising interactions  together,
interactions $M$ and $M'$ cancel and
  we  obtain the Ising representation 
  shown in Fig. 4.  The Ising model now has 
 five independent variables $J_1,J_2,J_3, J_4$ and $2P$.

\medskip
If we have further
\be
u_5=u_7,\hskip 1.5cm u_6=u_8,
\ee
then from the configurations  in Fig. 2, we see that the weights 
now possess an additional  up-down symmetry, namely,
\be
W(a,b,c,d) = W(d,c,a,b).
\ee
Consequently we have $P=-P$ implying $P=0$.  The Ising model representation 
is then of the form of a simple-quartic lattice with staggered interactions as shown in
Fig. 4 with $P=0$.

 If we have 
\be
u_5=u_6, \hskip 1.5cm  u_7=u_8,
\ee
 it can be seen from  Fig. 2 that
the $A$ weights have the symmetry
\be
W(a,b,c,d) = W(c,d,b,a)
\ee
and for $B$ sites we have
\be
W'(a,b,c,d) = W(-c,d -b,a).
\ee
In the resulting Ising model both $M$ and $P$ now cancel and the
lattice is shown in Fig. 5. 

\section{Summary and acknowledgment}
We have introduced an odd 8-vertex model for the simple-quartic lattice
and established its equivalence with a staggered 8-vertex model.
We showed that in the free-fermion case the odd 8-vertex model is completely
equivalent to the free-fermion model of Fan and Wu in a
noncritical regime.  Several Ising model representations of the
free-fermion odd 8-vertex model with pure 2-spin Ising interactions are also deduced.

\medskip
The work has been supported in part by NSF Grant DMR-9980440.
The authors would like to thank Professor Elliott H. Lieb for his aspiration
leading to this work. The assistance of W. T. Lu in preparing the figures is gratefully
acknowledged.

\newpage
\centerline{Figure Captions}
\bigskip
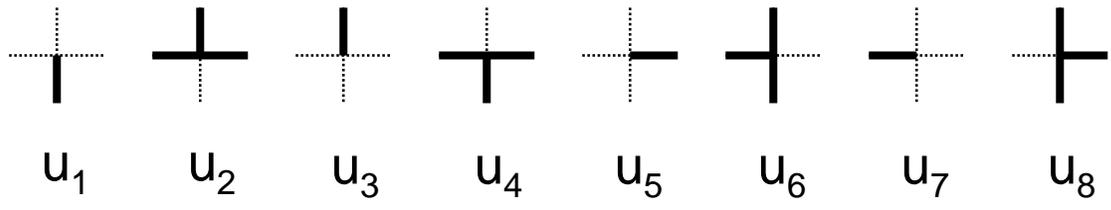
Fig.1. Vertex configurations of the odd 8-vertex model and the
associated weights.

\bigskip
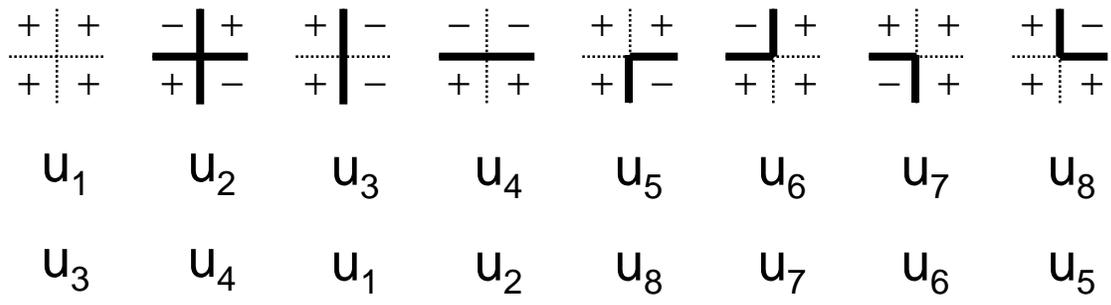
Fig. 2. An equivalent staggered 8-vertex model and the associated
spin configurations on the dual.
 
\bigskip
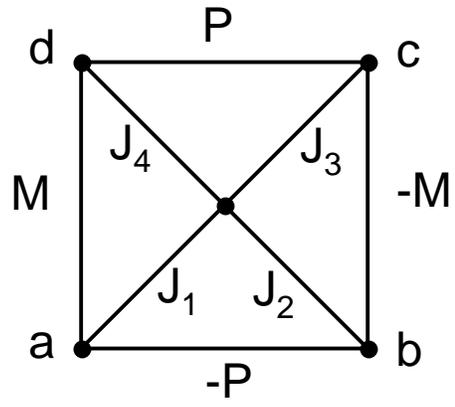
Fig. 3. Ising interactions in $W(a,b,c,d)$.

\bigskip
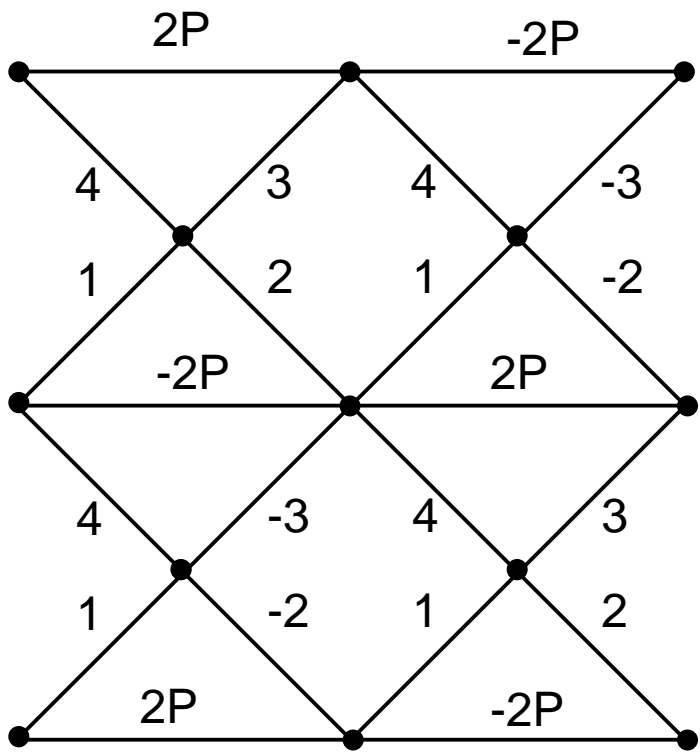
Fig. 4. An Ising model representation of the odd 8-vertex model.
The number $-2$ stands for $-J_2$, etc.

\bigskip
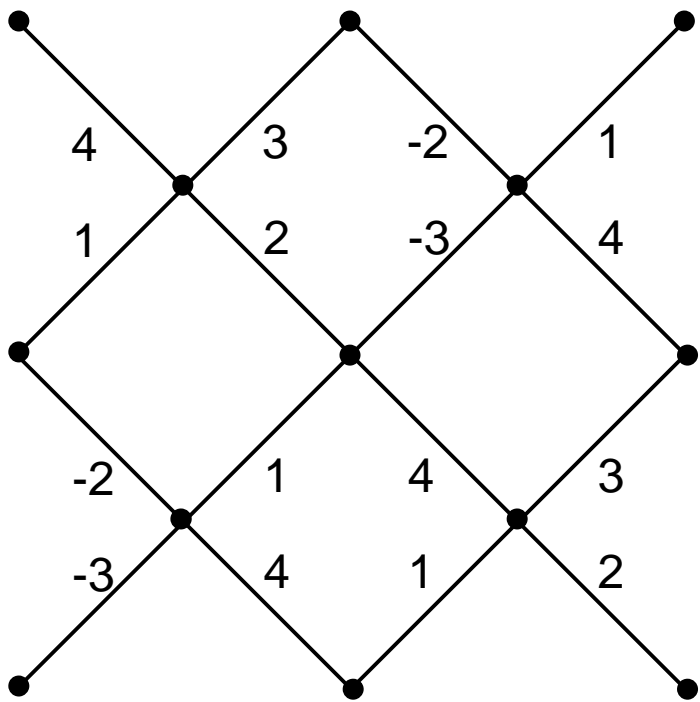
Fig. 5. An Ising model representation of the odd 8-vertex model when $u_5=u_6,u_7=u_8$.
The number $-2$ stands for $-J_2$, etc.

\end{document}